\documentclass[useAMS,usenatbib]{mn2e}
\usepackage{graphicx}
\usepackage{subfigure}
\usepackage{graphicx}
\usepackage{amsmath}
\usepackage{amssymb}
\title[BH sub-systems]{On black hole sub-systems in {idealized} nuclear star clusters}
\author[P. G. Breen and D. C. Heggie]{ Philip G. Breen$^1$\thanks{E-mail:
p.g.breen@sms.ed.ac.uk} and Douglas C. Heggie$^1$\thanks{E-mail:  d.c.heggie@ed.ac.uk} \\
$^1$ School of Mathematics and Maxwell Institute for Mathematical Sciences, University of Edinburgh, King's Buildings, Edinburgh EH9 3JZ}

\begin{document}



\maketitle

\label{firstpage}

\begin{abstract}
Recent observational evidence, numerical simulations and theoretical arguments seem to indicate that stellar mass black holes (BH) persist in stellar systems such as globular star clusters for much longer than previously thought.  Previously, theory implied that the BH would segregate into a compact system with short dynamical time scales, and that the BH would escape long before the present.  But stellar systems can exist in a state of {\sl balanced evolution}, where the energy generated in the core is regulated by the process of two-body relaxation in the bulk of the  system.  If the  system has a centrally concentrated BH subsystem and there is no massive central BH, then the energy is generated by three-body encounters in the core of this subsystem. Therefore, in balanced evolution, the evolution of the BH subsystem is regulated by the much longer time scales of the host system. In the present paper the implications of these results for { idealized} 
nuclear star clusters (NSC) are discussed. Though previous theory implied that BH would be almost absent from many NSC -- those with relatively short dynamical time scales -- it is argued here that, {based on the results of idealized models,} many such NSC could still be host to substantial BH subsystems.
\end{abstract}

\begin{keywords}
{ galaxies: nuclei -- galaxies: evolution -- galaxies: kinematics and dynamics}
\end{keywords}

\section{Introduction}\label{sec:intro}
The traditional view of the behaviour of stellar mass black holes (BH)
in  stellar systems such as globular clusters is as follows: the BH
rapidly segregate to the centre of the cluster, form binaries and
eject each other in super-elastic encounters, as argued by
\cite{Kulkarnietal1993} and \cite{Sigurdsson1993}. These arguments
imply that globular clusters should be almost completely devoid of
BH. This view has been significantly challenged recently from the
observational evidence of two { BH (perhaps implying a total
  population as large as 100)} in the globular cluster M22 \citep{Strader2012}. Numerical simulations have also indicated that BH could persist in  systems like globular clusters for longer then previously thought  \citep{Merritt2004,Mackey2007,Banerjee2010,Aarseth2012,hurleyShara2012,Morscher2012,Sippel2012}. A new theoretical treatment by \cite{breenheggie} also predicted longer time scales for the depletion of BH,  which they checked with numerical simulations {of idealized globular clusters}.

The treatment of  \cite{breenheggie} begins similarly to the traditional view, that is that most of the BH quickly segregate to the centre of the system. There the BH are concentrated together forming a subsystem of BH, which rapidly undergoes core collapse. The traditional view is that this subsystem then behaves as if it were independent from the host  system. 
However this is not the case; the BH subsystem sits in the deep
potential well of the other stars, and  there is an exchange of
kinetic energy between the BH and the other stars in the system. As
the other stars make up the bulk of the mass in the system, they
determine the flow of energy, and thus regulate the amount of energy supplied by the BH subsystem, and  hence its size. The energy supplied by the BH subsystem in turn has to be supplied by the core of the BH subsystem. This is done by the production and subsequent hardening of BH binaries.  When  these mechanisms  are in balance, ultimately it is the host system 
which regulates the production of energy in the core of the BH
subsystem. The escape of BH binaries and single BH is a byproduct of
the energy generation process, and is therefore  also regulated by the
host system. Hence the life time of the BH subsystem is determined by
the relaxation time scale in the much larger host system and not by
the much shorter timescale of the BH subsystem. Eventually,  however,  the BH subsystem does become seriously depleted, and then the system undergoes another core collapse until binary activity among the other stars becomes efficient enough, and balanced evolution is restored. 

NSC in galaxies have typical masses in the range $10^6$ to $10^7 M_{sun}$  \citep[][also see Fig \ref{fig:figone} in the present paper]{walcher2005}, which  is a factor $10$ to $10^2$ greater than the median globular cluster mass \citep[$8.1\times10^4$, see][]{HeggieHut2003}. NSC are very frequent, with around $75\%$ of galaxies containing them \citep{cote}. { NSC also have half-mass relaxation times larger than those of globular clusters \citep{merritt2009}.  The median half-mass relaxation time of the NSC in Fig \ref{fig:figone}  is $3.3 \times10^{9}$, assuming an average stellar mass of $1M_{sun}$.}  Nevertheless, as with globular star clusters, the random gravitational encounters 
between stars can substantially affect the structure of these systems, especially for more massive components and in the core. Ignoring the possible presence of a massive BH, NSC would be expected to evolve dynamically in a similar way to globular star clusters, that is, towards balanced evolution sustained by energy generation involving binaries. However \cite{MillerDavies2012} argued that the traditional mechanism of energy generation involving binaries would not be able to provide the energy necessary  if the central velocity dispersion exceeded $\sim 40km s^{-1}$ because in that case BH binaries would coallesce by emission of gravitational radiation before having time to interact dynamically with the other BH, and would then escape due to the kick applied at the final merger. They also argued that systems  whose central velocity dispersion exceeded that value would undergo a deep core collapse, possibly leading to the formation of a massive central BH.
 
In the present paper we will  reexamine 
the possibility of BH subsystems in NSC in the light of the above
recent advances in stellar dynamics.  We will  work with the aid  of
an idealized  NSC model,  consisting of two components (the BH and the
other stars), ignoring the dynamical role of the rest of the galaxy,
and assuming that the galaxy does not harbour a central massive BH. We
will assume as with globular star clusters that H\'{e}non's Principle
\citep{Henon1975} applies, i.e. that the energy generating rate of the
core is regulated by the bulk of the system. {Except for the issue
  of kicks resulting from merging BH binaries (see above) our
  discussion is newtonian.}  In the next section we will consider the
phases through which the evolution of the NSC proceeds. {Section 3
  considers  the evolution of an idealised NSC without dynamically
  active binaries, where the only mechanism of energy generation is
  escape of merged BH.} This is followed by {a section of} conclusions and discussion, {where various constraints on the existence of BH subsystems in NSC are discussed}.

\section{Evolution of BH subsystems in NSC}\label{sec:ev_nsc}

\begin{figure}
\includegraphics[scale=0.45]{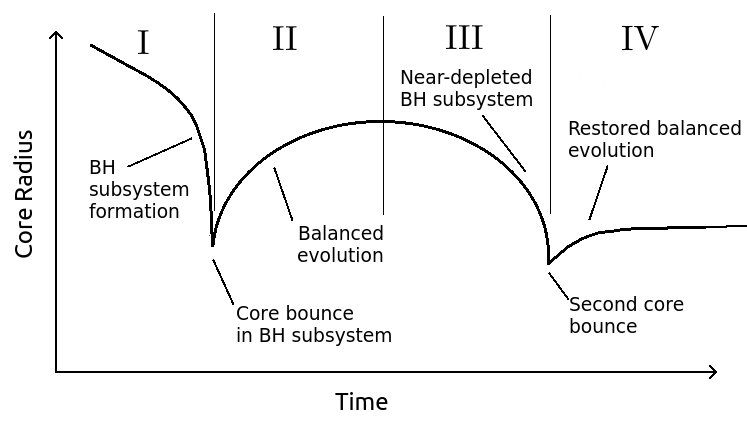}
\caption{{ Schematic} diagram of the evolution of the core radius in an idealized star cluster with a BH subsystem. 
The diagram is divided into four regions I, II,  III and IV. In region I the  BH segregate to the core forming the BH subsystem, which rapidly undergoes core collapse. This is followed by region II, where the system exists in a state of balanced evolution,  i.e. the core of the BH subsystem adjusts so that the energy it produces, through three-body interactions, balances the flow of energy by relaxation in the bulk of the system. As the BH population becomes  depleted in region II, in region III the system undergoes another core collapse. Finally in region IV balanced evolution is restored, although now the core of the low-mass stars  needs to be more compact to allow the formation and interaction of binaries composed of the less massive stars.}
\label{fig:dia}
\end{figure}
 
\begin{figure}
\includegraphics[scale=0.65]{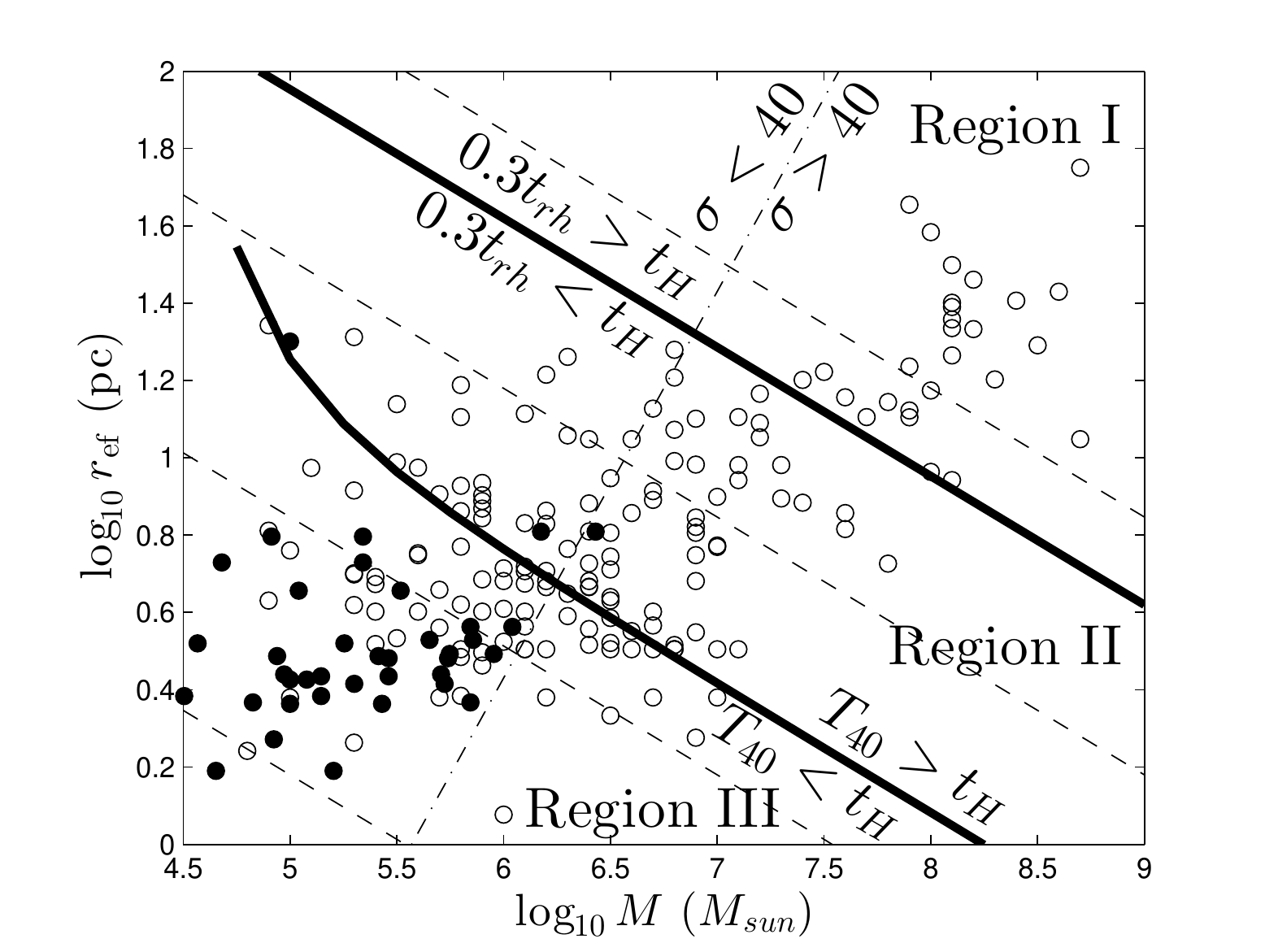}
\caption{ $r_{ef}$ and $M$ of a compilation of NSC \citep[unfilled
    circles, data from][]{sethetal2008} and a sample of globular star
  clusters \citep[filled circles, data from][]{markskroupa2010}. The
  dashed lines correspond to systems with $t_{rh}= 10^8, 10^9,
  10^{10}, 10^{11}$ { from bottom to top}. The upper solid line represents systems for which $0.3t_{rh}$ equals a Hubble time, and the lower solid curve represents those where the number of BH in the  system falls to about 40 ($T_{40}$, see text for assumptions) in a Hubble time. { This line is not present in the far left of the figure as the systems there are not massive enough to produce  40 BH under our assumptions}. The regions in the figure are summarised in the text and in Fig \ref{fig:dia} { which adds a fourth region}. Region II, between the two solid lines, is where systems containing a substantial centrally concentrated BH subsystem are expected to exist. See text for further details.}
\label{fig:figone}
\end{figure}

The evolution of the core of a system containing a BH subsystem is
illustrated in Fig \ref{fig:dia}, which is based on the ideas in
\citet{breenheggie}. The evolution has been divided into four regions:
region I, the segregation of the BH and formation of the dense central
BH subsystem; region II, balanced evolution powered by BH binaries;
region III, the depletion of the BH,  ending in a second core
collapse, this time in the low-mass component; and region IV,   the
ultimate  restoration of balanced evolution,  following the second
core bounce. The region a system will be in at the present time will
depend on its age and its evolutionary time scale, which in turn
depends on  the parameters of that system. In the idealistic case of a
system consisting of two populations, the BH and the other stars, these parameters are 
the half mass relaxation time, the stellar mass ratio of the two components and finally the total mass ratio of the two components.

For a star cluster with $N$ stars and a Kroupa IMF \citep{Kroupa} one would expect $\sim 2 \times 10^{-3}N$ BH to form \citep{Banerjee2010}.  Assuming that the BH are typically more { massive than other stars} by a factor $20$, then the total mass in BH is $\sim 4\%$ of the total cluster mass ($M$).  { To be conservative}, in the present paper we will assume that the total mass in BH is always $\sim 2\%$ of $M$ initially. We will also  make the simplifying assumption that the initial half mass relaxation time $(t_{rh,i})$ of a NSC is approximately the same as the present value $t_{rh}$. The effects 
of this assumption are discussed in Section \ref{sec:condis}.%

 Now we discuss the timescales of the various phases in Fig. \ref{fig:dia}.
A direct N-body simulation of an idealized two-component star cluster
with $N= 64k$ by \cite{breenheggie}\footnote{The parameters of the run
  were $N=64k$, { total mass ratio 0.02, and stellar mass ratio 20.}}
indicated that it takes $\approx 0.3t_{rh,i}$ for the bulk of the
stellar mass BH to {segregate to}  the centre of the cluster {and
  undergo core collapse}. In realistic systems the segregation time
scale for the BH depends on a number of factors, mainly the initial {spatial} distribution of BH and the stellar masses of the BH. {In an unpublished $N$-body simulation of the Galactic globular
  cluster M4, for example, core collapse of the BH subsystem took about
  $0.9t_{rh,i}$.}  Nevertheless we {shall} use the foregoing value as a guide, 
and assume that it takes approximately $0.3t_{rh}$ for the BH subsystem to form {and reach core bounce}.  Therefore systems which are less than $0.3t_{rh}$ old are still in the process of forming a BH subsystem, i.e. in region I of Fig \ref{fig:dia}. 

Shortly after this time balanced evolution is achieved, and the BH population {starts to become} steadily depleted (region II of Fig \ref{fig:dia}). The theoretical estimate from \cite{breenheggie} of the timescale of this phase of evolution is $\approx 6t_{rh,i}$. At some point there are insufficient BH remaining to maintain balanced evolution.  \cite{breenheggie} argued that {this happens} 
when the number of BH reaches about $40$. In the present paper, we
shall define the end of the subsystem as the time when the number of
BH reaches $40$ and assume the system then enters region III. Here the
system is no longer in balanced evolution and undergoes core collapse
{of the low-mass component}. If there are no BH remaining then the
core collapses until new binaries are formed from the other stars in
the system, and then balanced evolution is restored {(region IV in
  Fig \ref{fig:dia})}.   {If residual BH do remain, then
  contraction of the core of low-mass stars is still needed in order
  that interactions with the BH and with each other become efficient enough.}

Under our assumptions, there are two constraints {for the existence of a centrally concentrated subsystem of BH}: one, that the BH subsystem has had enough time to form (i.e. the age of the system is greater than $0.3t_{rh}$) and, two, that the BH subsystem has not been greatly depleted (i.e. the age of the system is less than $6t_{rh}$). There is also a lower limit to the total mass {of the system} of order $ 4 \times 10^4 M_{sun}$, as enough BH must form ($>40$) to evolve to a BH subsystem, as defined in the present paper.

The total masses and effective radii $r_{ef}$ (half light radii) are plotted in Fig. \ref{fig:figone} for a number of NSC \citep{sethetal2008}. The dashed lines in the plot represent lines of constant $t_{rh}$, which is defined as in \citet{Spitzer}, except that $r_{ef}$ has been used in place of the half mass radius. 
Our two constraints are represented by the solid black lines, {which means that} we can rule out the existence of dense central BH subsystems in regions I \& III {(which correspond to region I and regions III and IV, respectively, in Fig. 1)}.  However, it is important to remember that, in region I, centrally concentrated BH subsystems have not had time to form, though they could still be in the process of forming. Region II in Fig. \ref{fig:figone} is the region of interest in the present paper, where the NSC are still expected to contain BH subsystems (based on arguments and assumptions made so far.)

{Finally in this section} we will consider the critical central three dimensional velocity dispersion of $\sim 40km s^{-1}$ (see Introduction). This is represented as the dot-dash line in Fig. \ref{fig:figone}, assuming that $\sigma^2 \sim GM/r_{ef}$. \cite{MillerDavies2012} argued that above $\sim 40km s^{-1}$ the BH binaries would merge quickly as a result of emission of gravitational radiation and be promptly kicked out of the cluster. This would not allow the binaries to produce energy by the traditional mechanism of dynamical interactions involving binaries. However they neglected the energy generated by the escape of mass from the centre of the system. It can be shown from the theoretical arguments of \cite{Goodman1984} that, in a system of equal masses {in Newtonian dynamics}, the amount of energy generated by escaping stars/binaries is $\sim 50\%$ of the total energy generated, the rest being produced by encounters which result in an increase in the binding energy of the binary but do not result 
in generating 
escapers. 
As a result, we find that energy production, in systems in which BH
binaries escape without engaging in three-body interactions, appears
to be less efficient than in those in which three-body encounters have
time to act, but only by a factor of about 2  {for one-component
  systems. (Preliminary results from two-component systems indicate a
  factor in the range 2--5 depending on the stellar mass ratio; see Sec.\ref{sec:radclust}.)}{  {The effect of
    this is that the lower solid line in Fig.2 rises by an amount
    corresponding to a factor 2--5 increase in $t_{rh}$, in the region
    where $\sigma > 40$.}  Detailed evidence on this issue, including the case of unequal masses, is presented in Section \ref{sec:radclust}.} 

\section{Balanced Evolution without Binaries}\label{sec:radclust}

Since our theoretical discussion of Region II {in the
  Miller-Davies regime} is based on the notion of balanced evolution,
it is important to check that heating by loss of BH binaries {\sl
  without dynamical interaction} can adjust to produce the flow of
energy in the bulk of the system.  Indeed this mechanism is analogous to heating by loss of mass resulting from stellar evolution. It has long been known that mass loss from stellar evolution causes expansion in a star cluster \citep{Applegate}. However, it has recently been suggested that, under the right circumstances, a star cluster could exist in a state of balanced evolution powered by mass loss due to stellar evolution \citep{Gieles2013}. This is a similar  mechanism to the one suggested in the last paragraph of Sec.\ref{sec:ev_nsc}.

To check that stellar systems can exist in a state of balanced evolution without dynamically active binaries, but merely through loss of mass {as in the Miller-Davies regime}, we will now investigate the problem with an idealised model. One way in which this can be done is to modify a direct $N$-body integrator, in this case NBODY6 \citep{aarseth,NA2012}, to periodically check for binaries and give their centre of mass a large velocity kick. If this is done regularly enough and the velocity kick is large enough to remove the binaries from the system, 
any significant energy generation from dynamical interaction involving binaries is prevented. While this treatment is fairly crude, it is enough to test in principle if a star cluster can exist in balanced evolution without dynamically active binaries. We begin with the case of equal-mass systems, which allows us to test the concept of balanced evolution powered by the ejection of binaries in a simple and well understood case. 
Two-component systems will be considered more briefly below.

\begin{figure}
\subfigure{\scalebox{0.65}{\includegraphics[scale=1.]{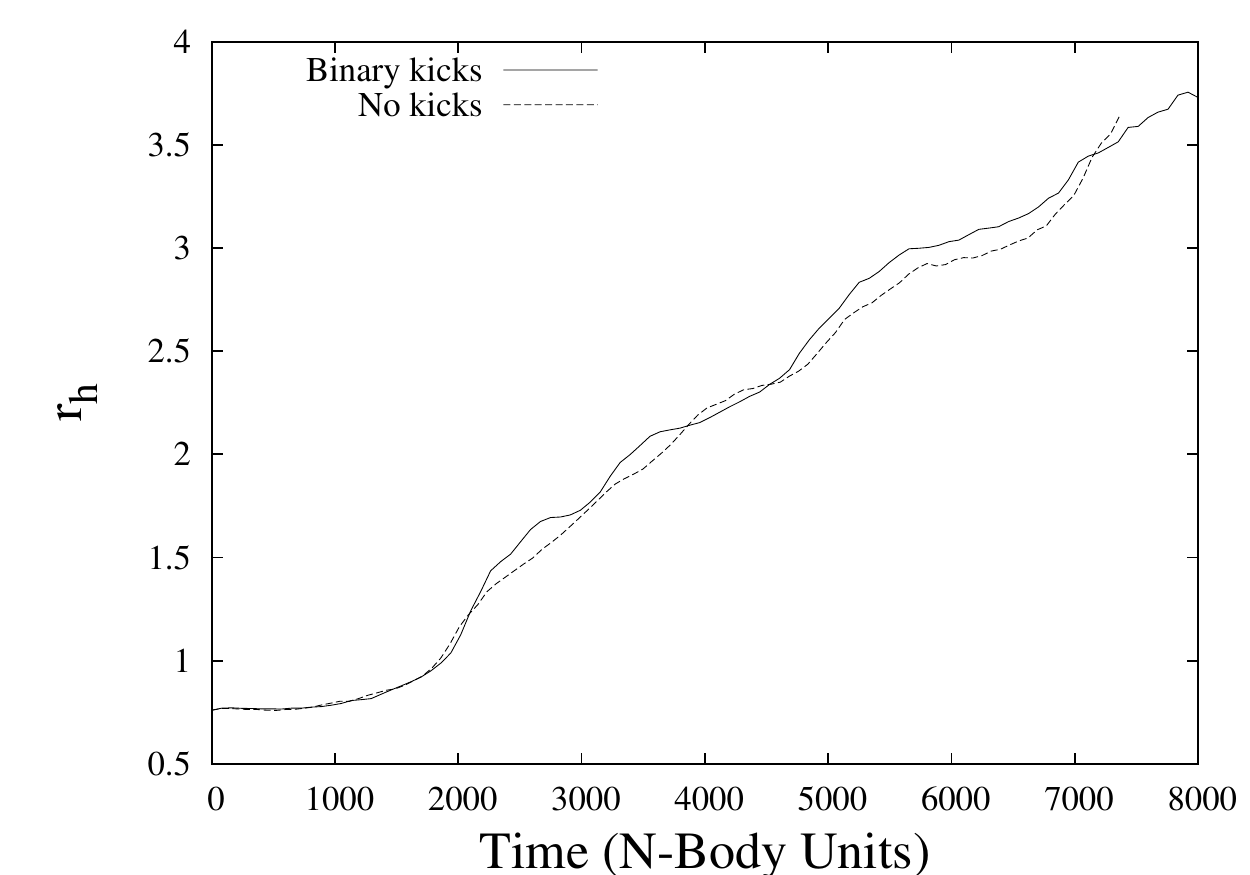}}}\quad
\subfigure{\scalebox{0.65}{\includegraphics[scale=1.]{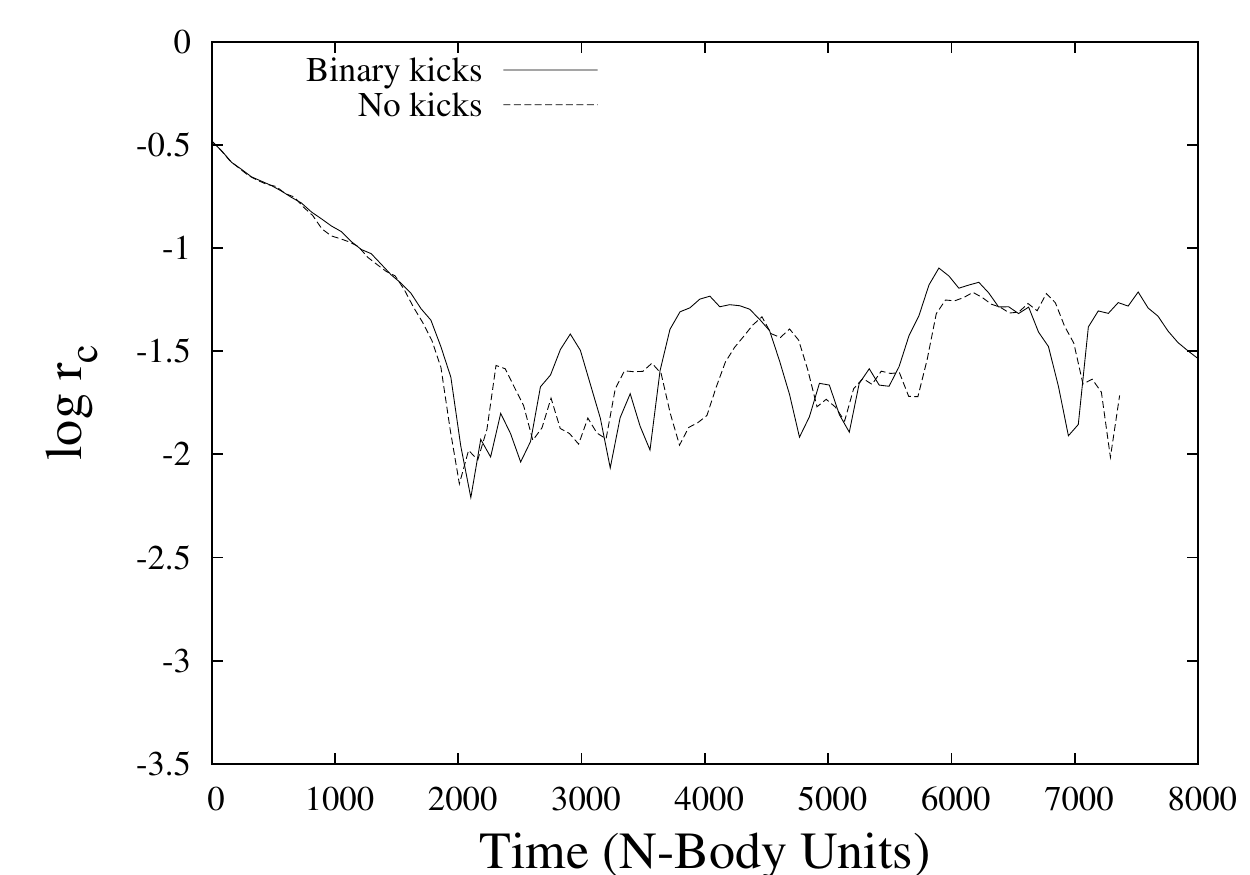}}}\quad
\caption{The evolution of the core radius ($r_c$) and half mass radius ($r_h$) {(N-body units)} of two one-component star clusters with $N=8k$ stars. In one of the runs, every $1$ N-Body time unit, all the binaries found in the system received a large velocity kick to their centre of mass (solid line). The kicks were large enough to ensure that those binaries escaped from the system. This prevented the binaries from having any significant dynamical interaction with the other stars. In the other run (dashed line) no velocity kicks were used and the binaries were allowed to interact dynamically with the other stars in the system. There is surprisingly little difference in the evolution of both $r_c$ and $r_h$, although there is a significant difference in the amount of high-energy escapers (see Fig. \ref{fig:NHE}).}
\label{fig:radclust}
\end{figure}

In Fig. \ref{fig:radclust}, two one-component N-body simulations are
compared, one which includes binary kicks and one which does not. As
can be seen in the figure, the evolution of $r_c$ and $r_h$ is
indistinguishable over the period of time shown, {except for fluctuations}. 

\begin{figure}
\includegraphics[scale=0.7]{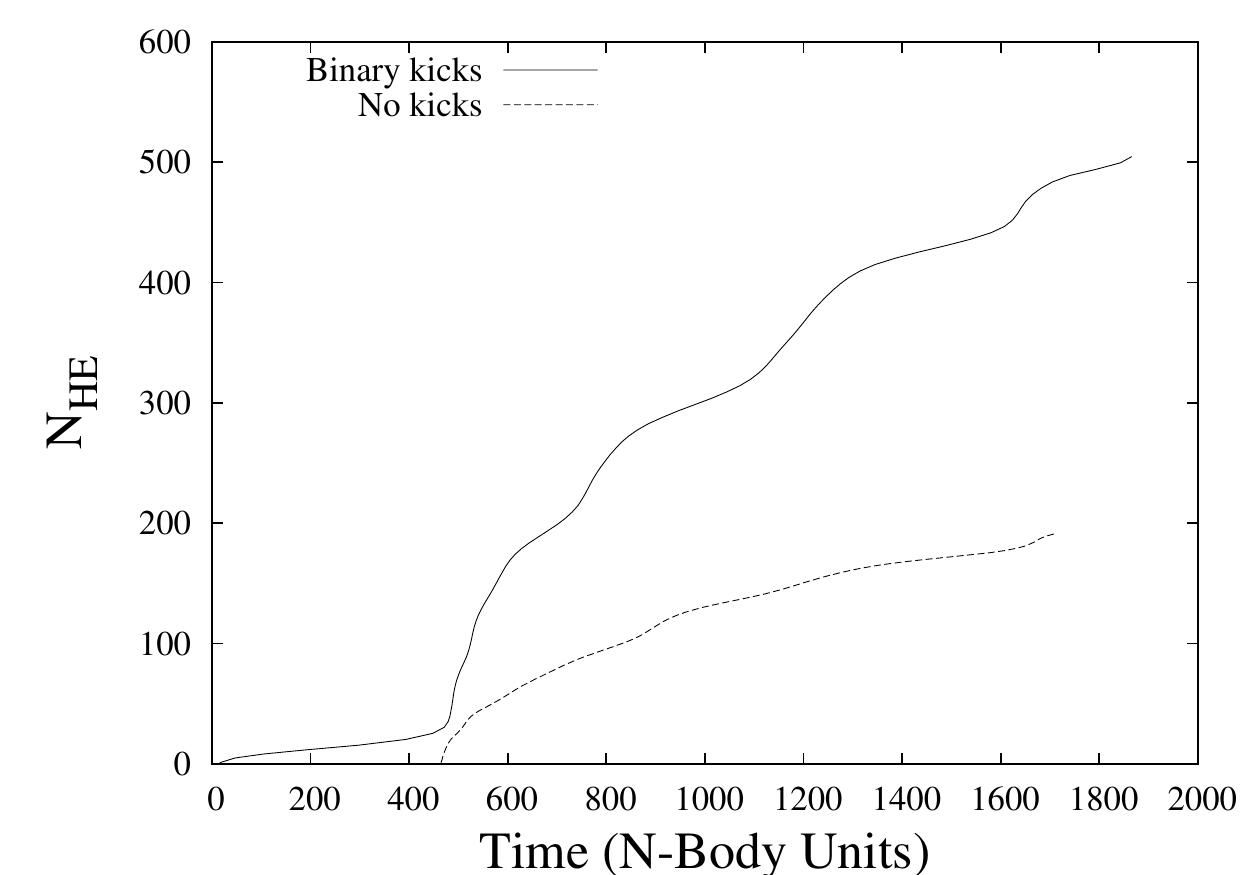}
\caption{Total number of high-energy escapers over time for the same runs as in Fig. \ref{fig:radclust}. A high-energy escaper is one which escapes with $v^2/\bar{v^2} >2$, where $v$ is the escaper's velocity ``at infinity'', and $\bar{v^2}$ is the average three dimensional squared velocity. The run which included binary kicks has a higher escape rate by a factor $\sim 2$, as expected  on theoretical grounds.}
\label{fig:NHE}
\end{figure}

Now we relate this behaviour to the 
mass loss in these systems. There are two distinct types of escapers in normal $N$-body simulations: those that result from two-body evaporation and those that result from energy-generating interactions in the core. These two types of escapers can be distinguished by the relative squared velocity \citep[i.e. $v^2/\bar{v^2}$; for example see][]{baum2002}; two-body evaporation produces escapers with  $  v^2/\bar{v^2} \lesssim 2$ and energy-generating encounters produce escapers with $v^2/\bar{v^2} \gtrsim 2$ (high-energy escapers). As both runs should produce similar numbers of escapers by two-body evaporation, we will ignore these and concentrate on the amount of high-energy escapers.  The number of high-energy escapers, $N_{HE}$, for both runs in Fig. \ref{fig:radclust} is plotted in Fig. \ref{fig:NHE}. As can be see from  Fig. \ref{fig:NHE} the high-energy escape rate is a factor of $\sim 2$ higher for the run with binary kicks. 

{On the assumption that the rate of energy generation is
  proportional to the rate of mass loss and the central potential,} this is what one would expect from the theory discussed in Sec.\ref{sec:ev_nsc}, as mass loss is the only energy source and this is only $\sim 50\%$ as efficient an energy source as compared with dynamically active binaries.  
 Though this establishes that mass loss can provide the energy required for ``normal'' cluster evolution, the fact that it is only half as efficient as energy production by three-body interactions does have an important consequence:  if applied to the case of systems containing a BH subsystem, it implies that the BH population would deplete itself in about half the time. This implies that  NSC in region II to the right of the dot-dash line depleted their BH population faster than we have allowed for and we can rule out systems in this area near the lower solid line.

 Now we consider briefly the application of these ideas, confirmed in
 the case of equal-mass systems, in the case of two-component systems,
 which we are using to model the evolution of clusters with a dense
 central BH subsystem. {The one-component systems in Fig
   \ref{fig:radclust} undergo a deep core collapse which increases the
   central potential significantly. In Fokker-Planck single-component
   models, for example, the ratio of the central potential to the mean square stellar speed more than doubles during core collapse \citep{Cohn1980}.  If a BH subsystem is present the host system does not undergo such a deep collapse and possibly may not collapse at all; though core collapse does occur in the heavy component, the heavy component contributes only of order 10\% of the central potential \citep{breenheggie}.   As the effectiveness of energy generation by the ejection of mass is proportional to the central potential, this energy source may be less efficient in a system containing a BH subsystem than 
in the one-component model show in Fig \ref{fig:radclust}. Empirical results from \cite{breenheggie} indicate that the amount of energy generation due to the ejection of BH may be as little as only $\sim 20\%$ of the total. This implies that the rate of mass loss in a BH subsystem is increased by a greater factor ($\sim 5$) than for the one-component models.  {Indeed} results of two $64k$ N-body simulations with the same initial conditions as used in \citet{breenheggie} indicate a greater increase in mass loss than seen in Fig. \ref{fig:NHE} { (in the sense that mass-loss of high-energy escapers is about 5 times higher in a system with binary kicks than in one with normal binary activity)}.}   {The upshot of this is that the loss of efficiency (if BH binaries cannot generate energy in three-body encounters) is greater than the factor of about two which we find for equal-mass models, and therefore the 
lifetime of the BH subsystem is also shortened by more than a factor of 2.}

\section{Conclusions and discussion}\label{sec:condis}

\subsection{Summary of conclusions}
{Based on an idealised two-component model,} we have considered the possibility of the presence of substantial dense central  BH  subsystems in NSC.  From our analysis it seems possible that a number of NSC could contain such  BH  subsystems. In Fig \ref{fig:figone}, we identified the region where dense central  BH  subsystems are most likely to exist (region II). The NSC in the other parts of Fig. \ref{fig:figone} have either not had enough time to allow for the segregation of the BH subsystem (region I) or have had sufficient time for the  BH  subsystem to be significantly depleted due to a phase of balanced evolution (region III). We have also shown in principle { (see Section \ref{sec:radclust})} that in the Miller-Davies regime, where the  BH merge too quickly to allow dynamical interaction with other BH, balanced evolution can still be achieved, but at the cost of a { higher  BH  escape rate and thus a} reduced life time.

\subsection{Formation of a supermassive BH}

\cite{MillerDavies2012} 
argued that the runaway merger of stars resulting from a deep core
collapse, which follows the depletion of the BH subsystem, results in
the formation of a central massive BH. From the results of the
idealized models  of \cite{breenheggie} it takes $\sim 10t_{rh}$  for
a system to reach the point of deep core collapse { (i.e. ``second
  core bounce'' in Fig.1)}.  { Therefore only systems below the {second} lowest dashed line in Fig.2 could form an
  intermediate-mass BH by this suggested mechanism.}  However the
energy generation {mechanism} in the models of \cite{breenheggie} is
more efficient than for the case of rapid binary ejection as a result
of merger by gravitational wave emission, { which tends to delay the
  onset of second core bounce} {in those models}.   Taking this into account, the implication is that only the NSC in region III (in Fig \ref{fig:figone}) bounded by the lower solid curve and the dot dash line, { might} 
be able to create a central massive BH by the mechanism suggested by \cite{MillerDavies2012}, unless either the evolution of the systems is considerably faster than  is estimated in the present paper, or the systems were initially much more compact, and have expanded into their present position (some consequences of which are discussed in Section \ref{sec:assumptions}).  

\subsection{Evaporation}

It is sometimes stated that the two-body evaporation time scale (i.e. a few hundred $t_{rh,2}$, where $t_{rh,2}$ is the {tiny} half-mass relaxation time {\sl of the BH subsystem}) places a strong constraint on the existence of a system of BH \citep[e.g.][]{Maoz}. This is reasonable if one is considering an isolated system, but if a system of BH is embedded in a much larger system (e.g. a NSC) evaporation can be substantially harder. This is because most of the BH which evaporate from the BH subsystem do { not} have high enough velocities to escape from the host system, but return to the BH subsystem on the mass segregation time scale. \cite{breenheggie} found for a model they considered that mass loss through evaporation was negligible, and that it was the effect of energy generation in the core which was the dominant cause of mass loss. Therefore 
the evaporation timescale { is not a relevant}  constraint on the
existence of BH subsystems {in the current context}.

\subsection{Simplifying assumptions}\label{sec:assumptions}
It is important to note that we are basing our conclusions on the
results of idealized models and we have ignored, for example, the
effects of stellar evolution, { or the interaction of the NSC with the
  rest of its host galaxy \citep{merritt2009}}. Some of these issues
are briefly discussed in \citet[see section 6.2]{breenheggie}.
One of the most relevant issues in more
realistic systems is the effect of a massive central BH. A massive
central BH could act as an alternative energy source, where energy is
generated by the binding of stellar mass BH to the the central massive
BH. If this were happening there would be a reduced need for energy
generation by stellar mass BH binaries, which ultimately reduces the
escape rate of stellar mass BH. This could potentially greatly prolong
the time scale of depletion of the  BH  subsystem.  {On the other hand
there is a sizeable class of galaxy where a central BH, if present, is
no more massive than the intermediate-mass BH claimed in some globular
clusters \citep{NW2012}.}

 Now we consider some of the approximations and simplifications which we have made in the course of this paper, and outline how some of these affect our results. Throughout this paper we have assumed that $t_{rh,i}\approx t_{rh}$, which is an approximation for two reasons. One reason is that a stellar system expands as energy is generated. This results in an increase of $t_{rh}$ with time.  Using the theory of \cite{breenheggie}, in a system with an initial BH population of $\approx 0.02M$ it is expected that $t_{rh}$ increases by a factor $\approx 3$ over the life time of the BH subsystem. Assuming the total mass loss is negligible, the systems in Fig \ref{fig:figone} evolve by moving upwards vertically. This means that systems in Fig  \ref{fig:figone} which started in region II could have expanded into region I (and similarly from region III to region II). Therefore it is  possible that some systems in region I contain a BH subsystem and are in the 
process of expanding. It is also  possible that region II contains systems which
originated in region III, but have depleted their BH population in the process of reaching their current position, or possibly have expanded as a result of other energy generation mechanisms (e.g. stellar evolution).   {The second reason why our time scales are approximate is that} we have also made the assumption that $r_h\approx r_{eff}$. {In reality} $r_{eff}$ is expected to be smaller due to projection effects { and mass segregation \citep{hurley2007}}. This implies that we are  {under}estimating the relaxation time (the solid { and dashed} lines should shift slightly downwards in Fig \ref{fig:figone}) and we are overestimating $\sigma$ (the dot-dash line should shift to the right). The latter implies that there are fewer systems in the Miller-Davies regime than Fig. \ref{fig:figone} indicates.

Finally in our discussion of simplifying assumptions that we have
made, we assumed that all BH that form in a stellar system do not
promptly escape as a result of natal kicks. The subject of natal kicks
for BH is still under debate \citep{Repettoetal2012}.  Nevertheless
natal kicks could reduce the initial BH population to a fraction of
the size that has been assumed in the present paper {(though we
  did make a conservative assumption on the initial number of BH, as
  mentioned in Sec.\ref{sec:ev_nsc})}. Given the estimates of \cite{Strader2012} of the present day BH population of M22 as between $5$ and $100$, however, it is not unrealistic to assume a significant initial BH population.

\subsection{Formation of a NSC}\label{sec:form_nsc}
NSC could have undergone a number of merger events, which change both $M$ and $r_{ef}$, either with other NSC as their host galaxy undergoes mergers, or with globular clusters dragged to the centre via dynamical friction. If the mergers happen rapidly enough our approximations are still appropriate for the subsequent evolution.  \cite{Cap} simulated a sample of four massive globular clusters in the inner region of a triaxial galaxy and found that they merged in much less than a Hubble time. However the merger process could also reduce the mass segregation time scale, as smaller systems undergo mass segregation more quickly,  and once they merge the mass segregation can be inherited from the smaller systems \citep{Vesperini2012}. The implication is, again, that some NSC in Region I may have already formed a BH subsystem.

\section*{Acknowledgment}
{We thank the referee for his comments, especially for pressing us to clarify the
  distinction between real NSC and our idealised models.}

\bsp

\newpage

\appendix

\label{lastpage}

\begin{thebibliography}{99}
\bibitem[\protect\citeauthoryear{Aarseth}{2003}]{aarseth} Aarseth S.J., 2003, Gravitational N-body simulations.  Cambridge Univ. Press, Cambridge
\bibitem[\protect\citeauthoryear{Aarseth}{2012}]{Aarseth2012} Aarseth S., 2012, MNRAS, 422, 841
\bibitem[\protect\citeauthoryear{Applegate}{1986}]{Applegate} Applegate, J.H., ApJ, 301 132
\bibitem[\protect\citeauthoryear{Banerjee et al}{2010}]{Banerjee2010} Banerjee S., Baumgardt H., Kroupa P., 2010, MNRAS, 402, 371
\bibitem[\protect\citeauthoryear{Baumgardt et al}{2002}]{baum2002}Baumgardt H., Hut P., Heggie D. C., 2002, MNRAS, 336, 1069
\bibitem[\protect\citeauthoryear{Breen \& Heggie}{2013}]{breenheggie}
  Breen P.G., Heggie D.C., 2013, MNRAS, 432, 2779 
\bibitem[\protect\citeauthoryear{Capuzzo-Dolcetta \& Miocchi}{2008}]{Cap}Capuzzo-Dolcetta R., Miocchi P., 2008, ApJ, 681, 1136
\bibitem[\protect\citeauthoryear{Cohn}{1980}]{Cohn1980} Cohn H., 
1980, ApJ, 242, 765 
\bibitem[\protect\citeauthoryear{C\^{o}te et al}{2006}]{cote}C\^{o}te P., Piatek S., Ferrarese L., Jordan A., Merritt D., Peng E. W., Hasegan M., Blakeslee J. P., Mei S., West M. J., Milosavljevic M., Tonry J. L. 2006, ApJS, 165, 57
\bibitem[\protect\citeauthoryear{Goodman}{1984}]{Goodman1984} Goodman J., 1984, ApJ, 280, 298
\bibitem[\protect\citeauthoryear{Gieles}{2013}]{Gieles2013} Gieles 
M., 2013, ASPC, 470, 339 
\bibitem[\protect\citeauthoryear{H\'{e}non}{1975}]{Henon1975}H\'{e}non M., 1975,  in Hayli A., ed., Proc. IAU Symp 69, Dynamics of Stellar Systems. Reidel, Dordrecht, p. 133
\bibitem[\protect\citeauthoryear{Heggie \& Hut}{2003}]{HeggieHut2003}Heggie D.C., Hut P., 2003, The Gravitational Million Body Problem. Cambridge Univ. Press, Cambridge
\bibitem[\protect\citeauthoryear{Hurley}{2007}]{hurley2007} Hurley 
J.~R., 2007, MNRAS, 379, 93
\bibitem[\protect\citeauthoryear{Hurley \& Shara}{2012}]{hurleyShara2012}Hurley J.,  Shara M., 2012, NMRAS, 425, 2872
\bibitem[\protect\citeauthoryear{Kulkarni, Hut \& McMillan}{1993}]{Kulkarnietal1993}Kulkarni S.R., Hut P., McMillan S.J., 1993, Nature, 364, 421  
\bibitem[\protect\citeauthoryear{Kroupa}{2001}]{Kroupa}Kroupa P., 2001, MNRAS, 322, 231
\bibitem[\protect\citeauthoryear{Mackey et al}{2008}]{Mackey2007}Mackey A.D., Wilkinson M.L., Davies M.B., Gilmore G.F., 2008, MNRAS, 386, 65
\bibitem[\protect\citeauthoryear{Marks \& Kroupa}{2010}]{markskroupa2010}Marks M., Kroupa P., 2010, MNRAS, 406, 2000
\bibitem[\protect\citeauthoryear{Maoz}{1998}]{Maoz}Maoz E., 1998, ApJ Lett., 494, 181
\bibitem[\protect\citeauthoryear{Merritt et al}{2004}]{Merritt2004}Merritt D., Piatek S., Portegies Zwart S., Hemsendorf M., 2004, ApJ, 608, L25
\bibitem[\protect\citeauthoryear{Merritt}{2009}]{merritt2009}Merritt, D. 2009, ApJ, 694, 959
\bibitem[\protect\citeauthoryear{Miller \& Davies}{2012}]{MillerDavies2012}Miller M. C., Davies M.B., 2012, ApJ, 755, 81 
\bibitem[\protect\citeauthoryear{Morscher et
    al}{2012}]{Morscher2012}Morscher M., Umbreit S., Farr W. M., Rasio
  F. A., 2012, ApJ, 763L, 15  
\bibitem[\protect\citeauthoryear{Neumayer 
\& Walcher}{2012}]{NW2012} { Neumayer N., Walcher C.~J., 2012, AdAst,
  2012, id. 709038}
\bibitem[\protect\citeauthoryear{Nitadori 
\& Aarseth}{2012}]{NA2012} { Nitadori K., Aarseth S.~J., 2012, MNRAS, 424, 545 }
\bibitem[\protect\citeauthoryear{Repetto et al}{2012}]{Repettoetal2012} Repetto S, Davies M.B., Sigurdsson S., 2012, MNRAS
\bibitem[\protect\citeauthoryear{Seth et al}{2008}]{sethetal2008} Seth A., Agueros M., Lee D., Basu-Zych A., 2008 ApJ, 628, 137
\bibitem[\protect\citeauthoryear{Sigurdsson \& Hernquist}{1993}]{Sigurdsson1993} Sigurdsson S., Hernquist L., 1993, Nature, 364, 423
\bibitem[\protect\citeauthoryear{Sippel \& Hurley}{2012}]{Sippel2012}Sippel A., Hurley J., 2012, MNRAS, 430, 30,
\bibitem[\protect\citeauthoryear{Spitzer}{1987}]{Spitzer}Spitzer L., 1987, Dynamical Evolution  of Globular Clusters. Princeton Univ. Press, Princeton, NJ 
\bibitem[\protect\citeauthoryear{Strader et al}{2012}]{Strader2012} Strader J., Chomiuk L., Maccarone T., Miller-Jones J.,
Seth A., 2012, Nature, 490, 71
\bibitem[\protect\citeauthoryear{Vesperini et al}{2007}]{Vesperini2012}Vesperini, E., McMillan, S. L. W., Portegies Zwart, S. F. 2009, ApJ, 655, 45
\bibitem[\protect\citeauthoryear{Walcher et al}{2005}]{walcher2005}Walcher C. J., van der Marel R. P., McLaughlin D., et al. 2005, ApJ, 618, 237
\end{thebibliography}
\end{document}